# Challenges and Solutions for Utilizing Earth Observations in the "Big Data" era

**Lachezar Filchev**  lachezarhf@space.bas.bg
*Space Research and Technology Institute,
Bulgarian Academy of Sciences - SRTI-BAS,
URL: http://www.space.bas.bg/Eng/Eng.html*

**Lyubka Pashova**  lpashova.niggg@gmail.com
*National Institute of Geophysics, Geodesy and
Geography, the Bulgarian Academy of
Sciences - NIGGG-BAS,
URL: http://www.niggg.bas.bg/en/*

**Vasil Kolev**  kolev_acad@abv.bg
*Institute of Information and Communication
Technologies, the Bulgarian Academy of
Sciences - IICT-BAS,
URL: http://www.iict.bas.bg/EN/*

**Stuart Frye**  stuart.w.frye@nasa.gov
*National Aeronautics and Space
Administration – NASA,
URL: https://www.nasa.gov/*

## Abstract

The ever-growing need of data preservation and their systematic analysis contributing to sustainable development of the society spurred in the past decade, numerous Big Data projects and initiatives are focusing on the Earth Observation (EO). The number of Big Data EO applications has grown extremely worldwide almost simultaneously with other scientific and technological areas of the human knowledge due to the revolutionary technological progress in the space and information technology sciences. The substantial contribution to this development are the space programs of the renowned space agencies, such as NASA, ESA, Roskosmos, JAXA, DLR, INPE, ISRO, CNES etc. A snap-shot of the current Big Data sets from available satellite missions covering the Bulgarian territory is also presented. This short overview of the geoscience Big Data collection with a focus on EO will emphasize to the multiple Vs of EO in order to provide a snapshot on the current state-of-the-art in EO data preservation and manipulation. Main modern approaches for compressing, clustering and modelling EO in the geoinformation science for Big Data analysis, interpretation and visualization for a variety of applications are outlined. Special attention is paid to the contemporary EO data modelling and visualization systems.

## 1. Introduction

Among the modern applications of Big Data sets, their number in the Earth sciences' usage has grown extremely worldwide thanks to the development of industrial technologies and the space programs of the leading countries in space explorations. The precise and cutting-age Earth Observation (EO) data provided by geospatial technologies empowered the modern society to tackle with challenges under the continuous environmental and climate changes. In the digital age, the rapid growth of processing power and global connectivity allows to collect, share and analyse the vast amount of EO data from geospatial surveying broadening their added-value.

Earth Observation applications are already facing the Big Data issue, with a need for advanced solutions supporting Big Data handling and Big Data analytics. There is well recognition that no single space program, agency or nation can hope to satisfy all of the observational requirements that are necessary for improved understanding of the Earth system. Geospatial Big Data sets are accumulated through the satellite images of over 660 satellites performing EO as listed by the Union of Concerned Scientists (UCS, April 2018). EOs together with new tools and technologies offer unprecedented opportunities to develop the capacities of countries to efficiently track all facets of the sustainable development (CEOS_EOBS, 2018).

There is a need for flexible solutions enabling ad hoc analytics on EO Big Data for scientific data exploration on demand. Processing, analysing and visualizing this data have its own challenges. All users require Big Data technologies supporting multiple data models and reducing data transfer as well as advanced visualization techniques easily integrated in different graphical user interface including Web and mobile systems. In addition to the achievements and the services provided until now, the new cloud-based Web services enables linking together multiple satellite capabilities into what is known as a satellite "sensor web" where additional insight into potential upcoming data acquisitions across a fleet of satellites can be obtained.

The benefits of EOs are already well understood across many areas of government, industry and science as a valuable information source in support of the social-economic development of modern society. Key benefits of satellite EO data for the achieving Sustainable Development Goals (SDGs) are following: i) they make the prospect of a Global Indicator Framework for the SDGs viable; ii) potentially allow more timely statistical outputs for reducing the frequency of surveys, respondent burden and other costs, and providing the data at a more disaggregated level for informed decision making; iii) contribute to improving the accuracy in reporting by ensuring that data are more spatially-explicit (CEOS_EOBS, 2018).

## 2. Mainstreams of EO Big Data Archives, Catalogs & Databases

Geospatial information is advancing in all the dimensions of Big Data. The high-resolution EOs have been constantly increasing in volume the last few years, and they are currently reaching petabytes in many satellite archives. The substantial advancement of space programs and Earth exploration of the renowned space agencies, such as NASA, ESA, Roskosmos, JAXA, DLR, INPE, ISRO, CNES, etc. contribute to expanding Big Data EO archives, catalogues, and databases. The international and national space agencies operate many research satellites for both EO and Space Science applications to validate the operational readiness of sensor and spacecraft platform improvements. It is estimated that up to 95% of the EO data present in existing archives has never been accessed, so the potential for increasing exploitation is a very big.

Satellite agencies are in the process of moving not only their data archive holdings to cloud-based storage, but are also transferring their processing and analysis software to the cloud services. For example, the United States National Aeronautics and Space Administration (NASA) is a research agency responsible for developing all civilian satellites for the U.S. government and provide free and open data from all U.S. civilian satellites at https://www.nasa.gov. NASA Earth Observing Systems (EOS) program has undertaken a project to migrate their 20+ year archive of satellite data products to the Amazon Web Service (AWS). In addition, they are implementing their processing software within cloud-based service offerings under configuration control of the Convex architectural tools. The EOS Project Science Office at NASA Goddard Space Flight Center serves to provide leadership in organizing open data, convening partners, and demonstrating the power of Big Data analytics through inspiring projects.

Copernicus programme, marking in 2018 already 20 years in development, has started in 1998 with the manifest of Baveno. Going through the semi-operational phase it was soon realised that the former Global Monitoring for Environment and Security (GMES) needs an easily recognisable name which turned out to be the Copernicus who lead the revolution of our understanding of the Solar system. Its pivotal role in astronomy is symbolic also for the remote sensing community, i.e. on how we look at our Earth as a system. It is Copernicus programme that brought the big data into practice with the introduction of Sentinels – the European flagship satellite constellation of Copernicus. Although not yet fully deployed the Sentinels already imposed an issue with data handling and manipulation. There are already national mirror sites which store satellite data from Copernicus rolling archive with an effort to provide a back-up solution for their regions as well as to make the Sentinel data more readily available to the local users.

Other national space agencies or corporations like Roskosmos, Japan Aerospace Exploration Agency, German Aerospace Center DLR, Brazilian National Institute for Space Research, Indian Space Research Organisation, French government space agency CNES and others support government's overall aerospace development and utilization, conduct integrated operations from basic space research and development, contribute to the Earth sciences, promote the practical use of EO data, and develop the technology of satellites, sensors and ground systems. For instance, the Chinese Government supports Big Earth Data Project starting in 2018, which will integrate science and technology infrastructure with research into resources, environment, biodiversity, and ecosystems. The project strives to make breakthroughs in Earth system sciences, life sciences, and associated disciplines. Most of the national space agencies produce high quality of science and technology in the space and terrestrial environment areas, provide unique remote sensing products and geoinformation services to the broader market, and managing large volume EO data for more than several decades. With the digital revolution and the fundamental shift towards digital and new 'Big Data' technologies, the issues at stake go well beyond the space sector and concern the construction of a whole new digital geo-information ecosystem. Following these new trends of innovative products developing, some of space agencies and companies offer fundamentally new services based on machine learning, integration with related services and technologies (navigation, geolocation, Internet of things, Big Data, etc.).

Recently, the Earth Observation industry sees six key technology drivers, i.e. Big Data Analytics, Cloud Computing, Artificial Intelligence, IoT, Automation and AR/VR transforming the way forward. Big Data and Cloud continue to be the two dominant technologies driving the geospatial industry (GeoBuiz, 2018). Satellite agencies are in the process of moving not only their data archive holdings to cloud-based storage, but are also transferring their processing and analysis software to the cloud. Inherent in cloud computing services are software tools that supply data analytics and data mining capabilities for users. Another feature that is being implemented on cloud-based platforms is known as the "Data Cube" where various satellite archive holdings can be ingested and multiple views of an entire area can be made comparing pixel to pixel of the same geolocation on the earth across many sensor types and across multiple years of archive holdings. Big Earth Data-Cube infrastructures are becoming more and more popular to provide Analysis Ready Data. Currently there is a wider effort to adopt a common definition of the Big Earth Data-Cube with the final goal of enabling and facilitating interoperability. Such "Data Cubes" are already in operation or under development as CEOS DC, Digital Earth Aust., Colombia DC, Swiss DC, etc. (Woodcock et al., 2018).

COPERNICUS DIAS's are a new way of the European Comission to provide an added value of the Sentinel satellite data through online computation services of five DIAS – Data and Information Access Services. These are currently: CREODIAS, SOBLOO, MUNDI, WEkEO, and ONDA DIAS's. They are developed through different consortia and ideally provide users to discover, manipulate, process and download Copernicus data and information. All DIAS platforms provide access to Copernicus Sentinel data, as well as to the information products from Copernicus' six operational services, together with cloud-based tools (open source and/or on a pay-per-use basis). For a smaller project with a research and development focus the Copernicus User Uptake is facilitated through the RUS – Research and User Support which provides the end user with a non-commercial research to use Sentinel and Copernicus data and products at large on online virtualised environments enabled with EO software necessary for scientific applications. It is RUS which is expected to facilitate the research community to uptake more actively Copernicus data – which is already limited through the leveraged hardware and software requirements. To do so, RUS provides also on site demonstration and hands-on trainings on particular topics of the end-user interest.

### 3. EO Big Data Analytics

EO Big Data is both a challenge and an opportunity. Big Data is "extensive datasets — primarily in the characteristics of volume, variety, velocity, and/or variability — that require a scalable technology for efficient storage, manipulation, management, and analysis." (ISO/IEC CD2 20546). For remote sensing Big Data, they could be more concretely extended to characteristics of multi-source, multi-scale, high-dimensional, dynamic-state, isomer, and nonlinear characteristics (Peng et al. 2018). New calculation methods, algorithms, research infrastructures and computational resources are highly demanded in order to handle the Big Data sets from EOs more efficiently for numerous real- and near-time applications. As the V's are increasing with times some scholars started to mock the V's concept by presenting the 42 V's The 42 V's of Big Data and Data Science concept alluring to the Universal number 42 as the ultimate answer (Shafer, 2017).

For instance the data volume of some of the most well-known EO satellite systems operated by public institutions is: Landsat 1-6 missions is 120 TB; The data volume of Landsat 7 and 8 is estimated from the relating metadata files provided by USGS; MODIS Terra and Aqua generate 70GB/day each; The data volume of Sentinel-1, -2 and -3, in their twin constellations is approximately 3.6 TB/day, 1.6 TB/day and 0.6 TB/day respectively (Soille et al., 2016).

The data flow from satellite, airborne and in-situ remote sensing is dramatically increasing and needs to be processing thanks to the "new" Big Data tools including with a new combination of former classical approaches. Advanced Big Data techniques are useful for preservation, compressing, clustering and modelling EO data during analysis, interpretation and visualization in a variety of applications. An important issue for EO Big Data manipulation concerns data standardization for enhancing their usability and interoperability. Standardized methods have been developed in order to assess, describe and propagate quality characteristics both quantitatively and qualitatively. For spatio-temporal 'Big Data', the OGC has defined its unified coverage model (nicknamed GMLCOV) which refines the abstract model of ISO 19123 (ISO 2005) to a concrete, interoperable model that can be conformance tested down to single pixel level. ISO and OGC are jointly developed the imagery and gridded data Reference Model, Framework and OGC Sensor Markup Language, Geography Markup Language (GML), and other ISO Metadata Profile.

Many useful algorithms and techniques for EO Big Data processing have been used recently, coming from the Internet of Things or from image mining in big EO image repositories. Cloud-based web object storage of EO data can simplify the execution of increased daily volume of satellite imagery. More importantly, it can facilitate EO data analysis of those volumes by making the data available to the massively parallel computing power in the cloud. However, storing EO data in cloud-based web object storage has a ripple effect throughout the space agencies' archive systems with unexpected challenges and opportunities. With the increasing multi-temporal and multi-sensor data, the multi-temporal classification of hipper-spectral images with high spatial and temporal resolution is happen a challenging problem.

Key issues in image fusion are balances between the preservation of spectral characteristics and high spatial resolution of the satellite images. Good fusion methods have to guarantee the preservation of the spectral information of the multispectral image when increasing its spatial information and must allow the injection into each band of the multispectral image. Many classical and advanced methods are applied for sub-sampling, resampling and multi-sensor EO data fusion such as Fourier transform, Maximum Likelihood, Artificial Neural Network, Decision Tree, Spectral Angle Mapper, Principal Component Analysis, Wavelet and Multiwavelet Transform (discrete and continuous).

Other useful technique is a deep learning. Before its applying, it is recommended to compress the EO data for improving the rapidity of the learning, but also for allowing an easy visualization by the user manipulating the raw information and by the data provider. The compression can be obtained by: (i) reducing the dimensionality of the observables (e.g., using Principal Component Analysis or Independent Component Analysis; (ii) discretizing (even binarizing) the data (e.g., using fuzzy or Fourier transform) before the steps of classification or data mining. In the last years, a pattern analysis algorithms as Support Vector Machines (SVM), Kernel Fisher Discriminant Analysis, and Kernel Principal Component Analysis can operate on general types of data and are powerful for classification and regression problems and general learning approaches. These methods are very popular tools as has been widely used in image denoising, image compression, image classification, image fusion and features extraction. Advantage method in processing HSIs is the multiresolution analysis (MRA) characterize by the prototype lowpass spatial filter that is adopted to design the specific bandpass decomposition leads to space - frequency localization at multiple resolutions and able to select features that can be used in the denoising technique and the reconstructed and/or make dimensionality reduction of HSI. Moreover, MRA is a unifying framework in which existing image fusion schemes can be accommodated and novel schemes devised and optimized [Alparone, 2015]. The MRA methods achieve good performance, temporal coherence, spectral consistency and acceptable computational complexity.

### 4. Conclusion

Many future satellite missions are envisaged until 2030 and beyond. The satellites take measurements of the Earth at various frequencies and at various electro-magnetic wavelengths including microwave, radar, lidar, and thermal-infrared, in addition to optical wavelengths in the 400-2400 nm range. The volume of EO is exponential increasing with advancement of sensor and digital technologies. The huge amount of data on the state of Earth planet and its changes has enabled the global monitoring of natural processes across the whole electromagnetic spectrum on an operational and sustained basis. Such streams of dynamic EO data offer new possibilities for scientists to advance our understanding of how the ocean, atmosphere, land and cryosphere operate and interact as part on an integrated Earth System. It also represents new opportunities for entrepreneurs to turn EO as Big Data into new types of information services. Some platforms for optimization the EO Big Data usage to accelerate the research efforts, and foster knowledge discovery and dissemination more quickly and efficiently for many added-value applications are developed based on High Performance Computing. However, these opportunities come with new challenges for scientists, businesses, data and software providers who must make sense of the vast and diverse amount of data by capitalizing on new technologies such as Big Data analytics (Mathieu and Aubrecht, 2018).

Recently scientists and academics have initiated an outreach programme "Geo for All"3 as a means of laying down strong foundations for Open Geospatial Science. Part of this initiative is to create openness in geo-education and to encourage the flair and creativity that is so critical to society's wellbeing, both now and in the future.